\documentclass{osa-article}

\journal{osajournal}

\usepackage{graphicx}
\usepackage[caption=false]{subfig}
 \usepackage{float}

\def\FORTB {far-off resonance optical dipole trap}
\def\MOT {magneto-optical trap}

\newcommand{\si}{$\sim$}
\newcommand{\um}{$\mu$m}
\newcommand{\uK}{$\mu$K}  
\newcommand{\us}{$\mu$s}

\newcommand{\Yb}{$^{171}$Yb}	

\newcommand{\clockTfull}{$(6s^{2})\,^{1}S_{0} -(6s6p)\,^{3}P_{0}$}

\articletype{Research Article}

\begin{document}


\title{Simulation of optical lattice trap loading from a cold atomic ensemble} %

\author{Raymon S. Watson and John J. McFerran}  %

\address{Department of Physics, University of Western Australia, 35 Stirling Highway, 6009 Crawley, Australia}  
\address{Corresponding author: john.mcferran@uwa.edu.au}  

\begin{abstract}
We model the efficiency of loading atoms of various species into a one dimensional optical lattice from a cold ensemble taking into account the initial  cloud temperature and size, the lattice laser properties affecting the trapping potential, and  atomic parameters.   Stochastic sampling and dynamical evolution are used to simulate the transfer, leading to estimates of transfer efficiency for varying trap depth and profile.  Tracing the motion of the atoms also enables the evaluation of the  equilibrium temperature and site occupancy  in the lattice.   The simulation  compares favourably  against a number of experimental results, and is  used to compute an optimum lattice-waist to cloud-radius ratio for a given optical power.  
\end{abstract}



\section{Introduction}

Optical lattice traps find applications within quantum simulation \cite{Dalibard2011Colloquium:Atoms, Tak2016},  tests of fundamental physics\cite{Wol2007,Zho2013}, and are an essential element of optical lattice clocks (OLCs)\cite{Ludlow2015, Nem2016, Gre2016a, Tyu2016}.  
Since their inception\cite{Katori2002}, OLCs have undergone rapid development leading to a
new era of relativistic geodesy\cite{Tak2016a,McG2018}, and have been used to perform one of the most stringent tests of general relativity through measurement of the gravitational redshift~\cite{Tak2020}.   The performance advantage of these clocks emerges from the coherent interrogation of a large ensemble of $N$ neutral atoms, with a signal-to-noise ratio which scales as $N^{1/2}$. 
 A larger value of $N$ generally  implies better stability and precision (collisional shifts aside).  Simulating the transfer of atoms from a  \MOT\ (MOT) to a lattice trap assists  with the design of experiments to produce large $N$. 
The localisation of atoms in the Lamb-Dicke regime~\cite{Neu1978,Jav1981b} is important in OLCs and related experiments~\cite{For2006b,Kar2009}, which motivates us to  focus on standing wave dipole traps\cite{Grimm2000} far from resonance.  
  Here the atom-light interaction is non dissipative to  a high degree,  thus, in the limit of far red (blue) detuning where scattering events by spontaneous emission can be neglected, we may treat the lattice trap as a purely classical potential where atoms are attracted to the local intensity maxima (minima) of the laser field. 
  Optical standing waves have been used to explore Bloch oscillations with ultra-cold atoms~\cite{Ben1996, Wil1996}, where the  
  the coherence length of the atoms extends across several periods of the potential.   For our study, although the atomic temperature is in the microkelvin range, it is  outside the regime of such quantum mechanical effects. 
%


Previously, modelling was performed on loading a \FORTB\ (FORT) from a MOT, where 1D atomic trajectories were considered\cite{Katori1999}.    In this case the FORT was produced with a travelling wave beam, where the stronger confinement is in the transverse direction, governed by the  beam profile. Careful consideration of the dynamics of FORT loading from a MOT has also been given in \cite{Kup2000}.  In the situation presented in the current work, the optical trap is produced with a focused standing wave beam,  
where the atoms are constrained to within half a wavelength in the axial direction and more weakly by the transverse profile.    
In work by Wu \textit{et al.}, optical lattice trapping  has been modelled, where it was assumed   
that all the atoms with kinetic energy less than the  potential depth were trapped~\cite{Wu2006}. 
 In our simulation we treat the ratio of lattice waist size to atomic cloud size as an adjustable parameter (relevant to most experiments), and compute the fraction of atoms remaining in the trap to investigate transfer efficiency. Adjustable parameters in the simulation include  trap depth,  atomic cloud temperature, cloud size, atomic mass,  laser waist size, optical power and atomic polarizability. 
We estimate the loaded fraction as a function of laser waist size for a given optical power and atomic cloud size, from which we determine the optimum trap-waist to cloud-radius ratio for a given laser power.
We compare our simulated trapping efficiencies  with known experimental results and find favourable agreement in most cases.   
We investigate the mean  occupancy per lattice site 
as a function of lattice depth and initial atom number, which may have significance for collisional shifts in lattice clocks. The Matlab code and User Documentation  for the simulation package are available in   the Supplementary Materials.

\section{Lattice potential and simulation}

Optical lattice trapping is often performed on dilute clouds of atoms that have been cooled to microkelvin  
 temperatures, for example, by the use of  magneto-optical traps (MOT). 
Here we simulate an ensemble of cold atoms, or atom cloud,  that is directly overlapped with the waist of a standing wave beam that forms a lattice trap  (in the majority of experiments  the atoms are located at the waist of the lattice beam).  The simulation involves full three dimensional kinematics and we assume the conditions of a far-off resonance trap, where excitations from the ground state are considered to be at very low rates in comparison to the lifetime of the trap. In this regime the dipole trap operates as a harmonic potential, where random absorptions 
can be neglected as they scale by the  inverse square of the frequency detuning, while the potential only scales as the inverse of the detuning\cite{Mil1993,Grimm2000}.

We assume that the MOT fields are switched off  during the transfer of atoms to the lattice trap, therefore  the atom cloud evolves with time in the    optical dipole trap potential without consideration of the MOT dynamics.   
For initialisation, the  atom cloud is generated by random sampling of the radial position of the atoms, in accordance with the  root-mean-square (rms) radius of the atomic cloud and  normal distribution (one class of Monte-Carlo method). These sampled radii (distance from  cloud center) are then assigned an isotropic distribution of unit vectors, which are multiplied by the  radii to achieve the appropriate 3D distribution of radial positions\cite{Knuth2002ComputerArt}.  
Thus formed is  a spherically symmetric distribution of atomic coordinates, which is valid under the assumption that the MOT is  sufficiently strong to counteract the gravitational force and  maintain a spherical shape (often not the case for narrow-line  MOTs). Each atom is then assigned velocity components $(v_x,v_y,v_z)$, where each component conforms to the one-dimensional Maxwell velocity distribution using the designated temperature of the cloud. Again we assume spherical symmetry for the velocity distribution, which is approximately valid in the regime of equilibrium  after cooling and trapping in a MOT (where the restoring force is  made the same in all six directions).

\begin{figure}[htp]
\begin{center}
{%
  \includegraphics[clip,width=0.5\columnwidth]{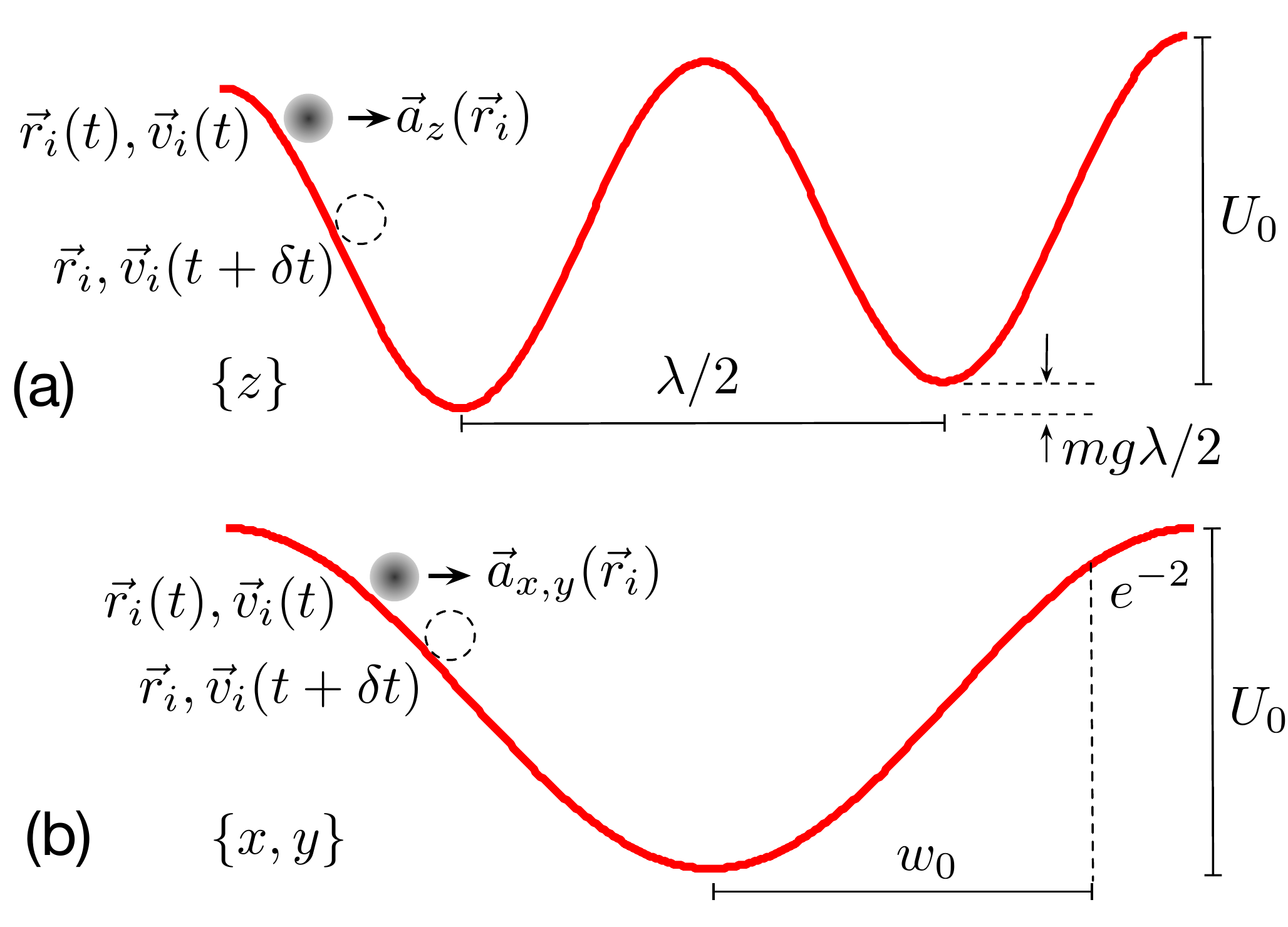}%
}

\caption{A sketch of the (a) axial and (b) traverse potentials and influence on the atomic trajectory. (a) Periodic axial potential   of depth $U_0$ situated every half wavelength giving rise to an acceleration $\vec{a}_z(\vec{r_i}$) as a function of position $\vec{r_i}$. The index $i$ labels each atom.  The simulation occurs over discrete time steps of duration $\delta t$.  (b) Transverse potential  due to the TEM$_{00}$ mode  of a gaussian-spherical beam profile with waist size  $w_0$.  
}
     \label{FigTrap}
\end{center}
\end{figure}

The lattice potential produced by a standing wave laser beam has the form~\cite{Sch2000c},  
\begin{equation}
    U(r,z) = U_0 ( 1 - \cos^2(k z) e^{- 2 r^2/w(z)^2 }    ) + mgz
    \label{Eq1}
\end{equation}
where $r=\sqrt{x^2+y^2}$ is the transverse radius and $w(z)$ is the laser spot size as a function of the axial position $z$, which varies as $w(z)^2 =w_0^2+(z\lambda/\pi w_0)^2$, where $\lambda$ is the wavelength of the trapping light and $w_0$  is the laser waist size ($e^{-2}$ radius).     A sketch of the potential  is shown in Fig.~\ref{FigTrap} for (a) longitudinal and (b) transverse directions.  
 The $\cos^2(kz)$ term forms the axial lattice provided by the standing wave, where $k$ is the wavenumber. The $mgz$ term is the gravitational potential in the $z$ direction, where we consider  the axial direction of the lattice trap to be aligned vertically (which impedes intersite tunneling), $m$ is the atomic mass and $g$ is the gravitational acceleration. 
 The term $U_0$  is the maximum energy level shift produced by the lattice beam, given by, 
 \begin{equation}
 U_0 = \frac{4\alpha(\lambda)P}{ \pi c \epsilon_0w_0^2}  
    \label{EqDepth}
\end{equation}
where $\alpha(\lambda)$ is the atomic ac  polarizability,  $c$ is  the speed of light, $\epsilon_0$ is the permittivity of free space, and $P$  is the unidirectional laser power. An appropriate scaling for the lattice depth is $U_0/k_B$, where $k_B$ is Boltzmann's constant. Another scaling is
with the single photon recoil energy $E_R = \hbar^2 k^2 / 2m$, where  $\hbar$ is the reduced Planck's constant.  
An advantage of the   former is that it  is not specific to an atomic species. We will make use of both.
For the atoms considered below, \{Yb, Hg, Sr\},  1\,\uK\ $\equiv\{10.3, 2.73, 6.08\}\, E_R$, respectively, where $k$ is set by the magic wavelength of the \clockTfull\  (clock) transition. 


 Evolution of the lattice trapping is accomplished through simulating the atomic trajectories over small discrete time steps $\delta t$.   The selection of a suitable time step is discussed below.  
Over each time-step the atomic trajectories evolve in a  manner according to,
\begin{equation}
\vec{r}_{i}(t + \delta t) = \vec{r}_{i}(t) + \delta t . \vec{v_i}(t)  
 \label{EqKin1}
\end{equation}
\begin{equation}
\vec{v}_{i}(t + \delta t) = \vec{v}_{i}(t) - \delta t . \vec{\nabla} U(\vec{r}_{i}) /m  
 \label{EqKin2}
\end{equation}
where $\vec{r}_i$ is the position coordinate and $\vec{v}_i$   the velocity  of atom number $i$. The acceleration is generated by the lattice potential through $-\vec{\nabla} U(\vec{r}_{i}) /m$.  The kinematic terms are illustrated in Fig.~\ref{FigTrap}.  To be captured in the standing wave  trap, the atoms must be within a certain range of initial positions  and with a low enough velocity to be maintained within the potential. The main means by which atoms exit the trap is through the weaker transverse confinement. Due to the larger well size in the radial direction, atoms have a much larger range of oscillatory motion. However, when their position  extends into the wings of the intensity profile, $\partial U/\partial r$ is significantly reduced, and
 the atoms leave the trap when the radial confinement loses out to  their transverse kinetic energy.

The ensemble evolution is  computed 
on a quad-core pc with use of the parallel computing package available in Matlab \cite{Mathworks1994Matlab:Toolbox}.   As individual atoms can be treated on separate cores it means each atomic evolution is treated independently, and there is insufficient  knowledge concerning the simultaneous location of atoms, thus we rely on the assumption that the number of collisions is insignificant. 
This is feasible   given the low atomic densities ($<10^{10}$\,cm$^{-3}$) and time scales  considered, but it  implies that 
we cannot model thermalisation through the 3 spatial dimensions.  
 In addition, the trap depth is assumed to be deep enough to  mitigate significant rates of atomic tunnelling~\cite{Lem2005} (i.e., greater than a few $E_R$).  A single step of the simulation 
 computes  the atomic positions and velocities for the given set of atom and trap parameters as a function of time. The simulation proceeds until there is negligible reduction in atom number $-$ equivalent to reaching equilibrium (the evolution time is set as an input parameter, further details are in the Supplementary Material).  
 A simulation sequence consists of  stepping the value of an input parameter and running the simulation  for each step; an example being the potential depth.   To obtain uncertainties on the output parameters, simulations are run approximately  ten times for a given parameter set and the standard deviation calculated.   An entire sequence of simulations may consist of more than 100 individual lattice trapping simulations.  Most simulations are carried out for an initial $10^4$ atoms (the absolute value is  not critical   when concerning  transfer fractions). 
 Simulations involving $10^6$ atoms or more would likely require a mainframe computer to carry out a single simulation within several hours.

Within the  simulation, the atomic mass, trap depth and profile, initial cloud temperature, time-step and ac polarizability are all adjustable parameters. Outputs from the simulation are principally  the trapped atom percentage, the final temperature, and the mean lattice site occupancy.  We discuss each of these below.

\section{Results and Analysis}

The optical dipole trap simulation is applicable to any conservative dipole trap with a wide array of possible parameters and trap types. The initial trap structure used here is based on the trapping of  $^{171}$Yb atoms with  light at wavelength  $\lambda=759.4$\,nm (magic wavelength for the \clockTfull\ transition~\cite{Lem2009a}).   An important parameter is the ratio  of  the waist  radius of the lattice beam, $w_0$,  to the   rms atomic cloud radius, $r_a$, denoted here as $\xi=w_0/r_a$. 
 The atomic cloud profile assumes the form of a gaussian, $f(r)= \exp(-r^2/2r_a^2)$.

 \begin{figure}[h!]
    \begin{center}
        \includegraphics[width=0.9\textwidth]{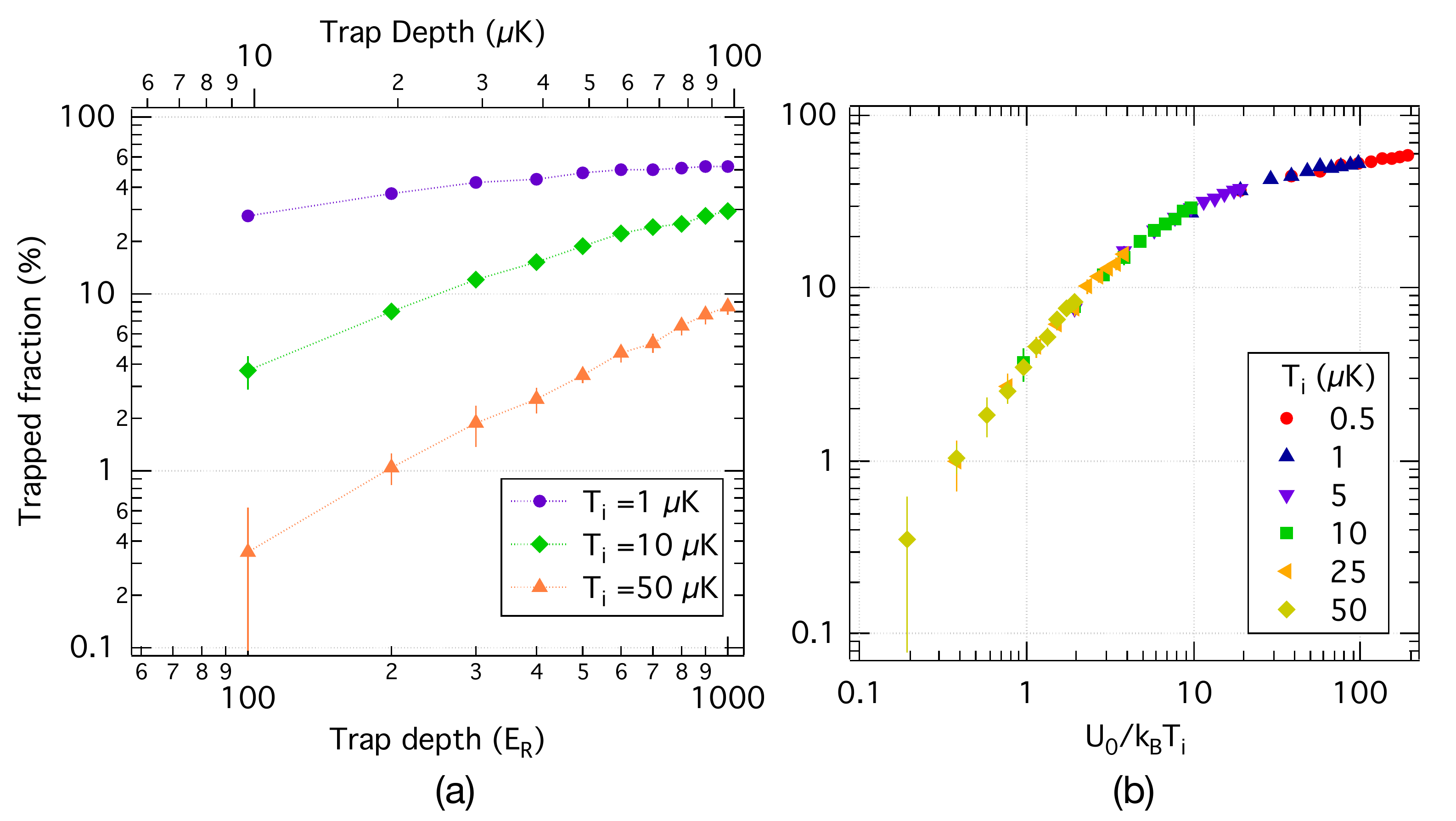}
        \caption{(a) Percentage of Yb atoms trapped  versus lattice depth for  different atomic cloud temperatures, $T_i$. 
         (b)  Percentage of Yb atoms trapped versus lattice depth  divided by the initial cloud temperature.  For both plots $\xi=0.5$. 
        }
                \label{Fig_fvsU}
    \end{center}
\end{figure}

\begin{figure}[h!]
    \begin{center}
        \includegraphics[width=0.9\textwidth]{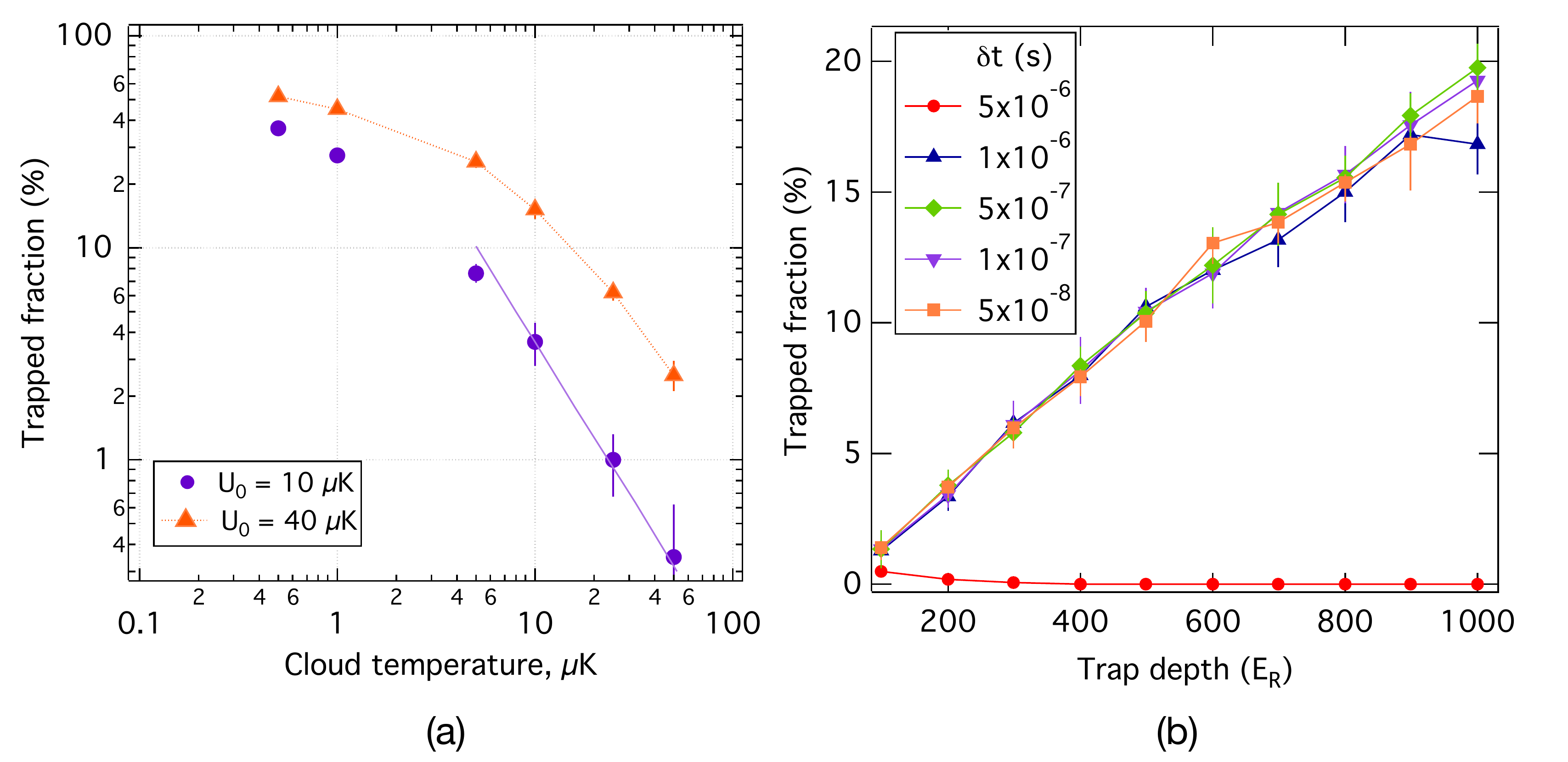}
        \caption{ %
        (a) Trapped fraction as a function of  atomic cloud temperature for  different lattice depths.   $\xi=0.5$.
        (b) Trapped fraction versus lattice depth for a range of time-steps.      
        } 
               \label{Fig_fvsT}
    \end{center}
\end{figure}

 Figure~\ref{Fig_fvsU}(a) shows  the    fraction  of atoms transferred, $f_t$, from the atomic cloud  to the lattice trap as a function of the trap depth, $U_0$,  for  different cloud temperatures, $T_i$.    Here, $\xi=0.5$  with  $w_0=85$\,\um\ and $r_a=170$\,\um\ (the dependence on $\xi$ is examined below).
  Higher transfer efficiencies occur for lower atomic temperature, in line with expectations.  However, the trapped fraction is seen to asymptotically approach a certain maximum value, 
     where the  trapping efficiency is determined predominantly by the fractional overlap of the laser beam and atom cloud, rather than beam intensity (i.e., at a depth where nearly all the atoms that were initially within the beam have been trapped, the loaded fraction can no longer increase with further increases in depth).  The error bars correspond  to the standard deviation of $f_t$  produced from \si\ 10 simulations of the same parameter set.    From previous computations of lattice trap loading~\cite{Wu2006}, one may expect to see a $f_t=aU_0^{3/2}$ scaling between trapped fraction and lattice depth (where $a$ is a constant).   At the lowest depths considered here, or high atomic cloud temperature,  we find $f_t=aU_0^{1.3}$.    The difference in the exponent is  because we  consider a  trap volume smaller than  the atom cloud.  
    While the computation here was specific to Yb, it may be applied to other atoms if the mass is taken into account (see below) and the trap depth in \uK\ is used (upper abscissa). 
Figure~\ref{Fig_fvsU}(b) presents a modified version  of Fig.~\ref{Fig_fvsU}(a), where the  lattice depth is  normalised by the initial cloud temperature, $k_B T_{i}$.  
 We see that the trapped percentage follows one continuous distribution.   Hence,  for a given beam width, the dominant factor in determining  
  transfer efficiencies is the dimensionless ratio of trap depth to the average thermal energy of the cloud.   This is in line with expectations, as an increased number of atoms will remain  untrapped if their initial energies are able to overcome the potential barrier provided by the lattice beam.  
  At the lowest depths the transfer fraction follows the $f_t\propto U_0^{3/2}$ scaling in accord with~\cite{Wu2006}. 
   Figure~\ref{Fig_fvsT}(a) shows  the  percentage   of atoms trapped as a function of temperature.  There is not a simple power law relationship between the two, but away from the geometric constraint   $f_t\propto T^{-3/2}$, as represented by the solid line.

Tests were carried out  to find an appropriate time step for simulation.   Influential is the time  an  atom takes to traverse a small distance  of a well under maximum acceleration.  In Fig.~\ref{Fig_fvsT}(b) we show the transfer fraction as a function of trap depth for a range of time steps (and for the same atomic and lattice parameters as mentioned above).    The results  show reproducible  output values  for $\delta t \leq 1 \times 10^{-6}$\,s.  Time-steps of $\delta t \geq 5 \times 10^{-6}$\,s  
produce a marked reduction
 in atom trapping percentage and are deemed invalid.  The disparity occurs because one is not tracking the atomic trajectories with sufficient resolution through the potential wells. 
For example, in the extreme case, atoms may  
  come in contact with a region of high acceleration at a well boundary, following which a large translation is made and the atom is no longer in the vicinity of the trap $-$ it misses experiencing the restoring force from other well boundaries.   We also see a 
  divergence when the potential depth increases beyond $U_0 = 900\, E_R$ with $\delta t = 1 \times 10^{-6}$\,s (deeper traps create greater accelerations). Thus, care should be taken when changing simulation parameters such that the time step remains valid.  In a further test we evaluated the final temperature for a range of time steps.  For the same parameters above, but with $U_0=10^3\, E_R$ and $T_i=10$\,\uK, the simulation produced the same final temperature (13.8\,\uK) with a standard deviation of 0.13\,\uK\ for ($0.05 <\delta t<3)$\,\us.  For the majority of the  computations reported here $\delta t = 0.1$\,\us.

\begin{figure}[H]  
    \begin{center}
        \includegraphics[width=0.9\textwidth]{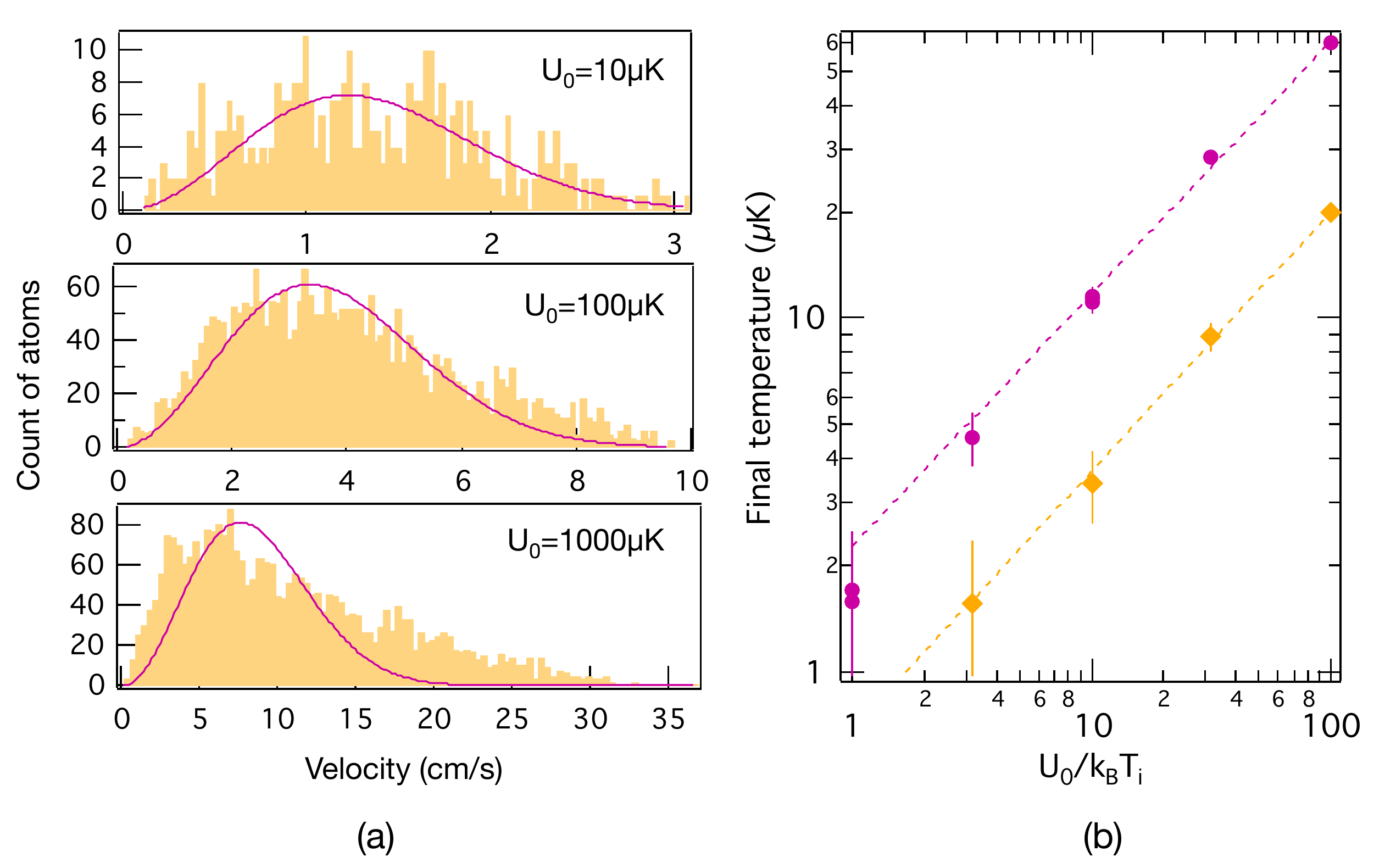}
        \caption{(a) Speed distributions for three lattice depths derived from the norm of the final velocity vectors.  The initial cloud temperature was 10\,\uK. 
        (b) Temperature of the atoms in the lattice trap as a function of trap depth, where the trap depth is normalised by the initial cloud temperature. The temperature is determined from all three components of $\vec{v}$ (circles), or from the two transverse components of $\vec{v}$ (diamonds). 
         }
          \label{Fig_Tf}
    \end{center}
\end{figure}

To evaluate the final temperature of the remaining trapped atoms, the Euclidean norm of the final velocities is taken and a histogram generated to which is fit  the Maxwell-Boltzmann distribution for speed, $f(v)\propto v^2\exp(-mv^2/2k_B T)$ (and available as a Matlab fitting function: \textit{fit\_maxwell\_pdf}\cite{Gal2004AFunctions}). 
 Examples of these distributions are shown in Fig.~\ref{Fig_Tf}(a) for lattice depths of 10\,\uK, 100\,\uK\ and 1000\,\uK\ (from top to bottom). Here the initial cloud temperature was 10\,\uK. The simulations were carried out with $10^4$ atoms and $\xi=0.5$. For the lower trap depths the distribution conforms reasonably well to the Maxwell-Boltzmann distribution, but when $U_0/k_B T_i\sim100$  there is a distinct deviation. We will return to this point momentarily. The relationship between  final   temperature and the depth of the lattice  is shown in Fig.~\ref{Fig_Tf}(b), where the trap depth is normalised by the initial cloud temperature  (i.e., $U_0/k_B T_i$).  There are two sets of data: circles represent the temperature derived from all three components of the $\vec{v}$,  and  diamonds where the temperature is derived from the transverse  components of $\vec{v}$.   In both cases there is a power law dependence with index \si~0.7. We note that for $U_0/k_B T_i=100$  the final 3D temperature (circles)  is about six times greater than the initial temperature of 10\,\uK. This has some support from experiment where a rise in temperature has been observed in a FORT~\cite{Fuk2007b}.  
  We may explain the rise as follows.
Since we assume an initial  normal spatial distribution of atoms, some low energy atoms can reside near the edge of a potential well.  As they fall into the well they gain kinetic energy, correspondingly, some portion of atoms experience an increase in velocity.  The deeper the trap, the greater the gain in KE.  In practice,  MOT and lattice fields usually operate simultaneously before the MOT fields are extinguished. In this case, one would expect some localisation in the wells before the MOT light is extinguished, with the increase in temperature being less apparent.  We note also that sideband cooling is often used to further reduce the temperature of atoms in a lattice trap~\cite{Ham1998,Ido2002}. 
By considering only the traverse components of the velocity one avoids the majority of instances where atoms reside near the maxima of the wells.  In this case the associated temperature is  a factor of three lower.   We also attribute the gain in KE of some atoms to the change in shape of the speed distribution seen in the bottom plot of Fig.~\ref{Fig_Tf}(a).  This deviation in distribution is considerably less if  only  the transverse velocity is considered.  

\begin{figure}[h!]
    \begin{center}
        \includegraphics[width=0.6\textwidth]{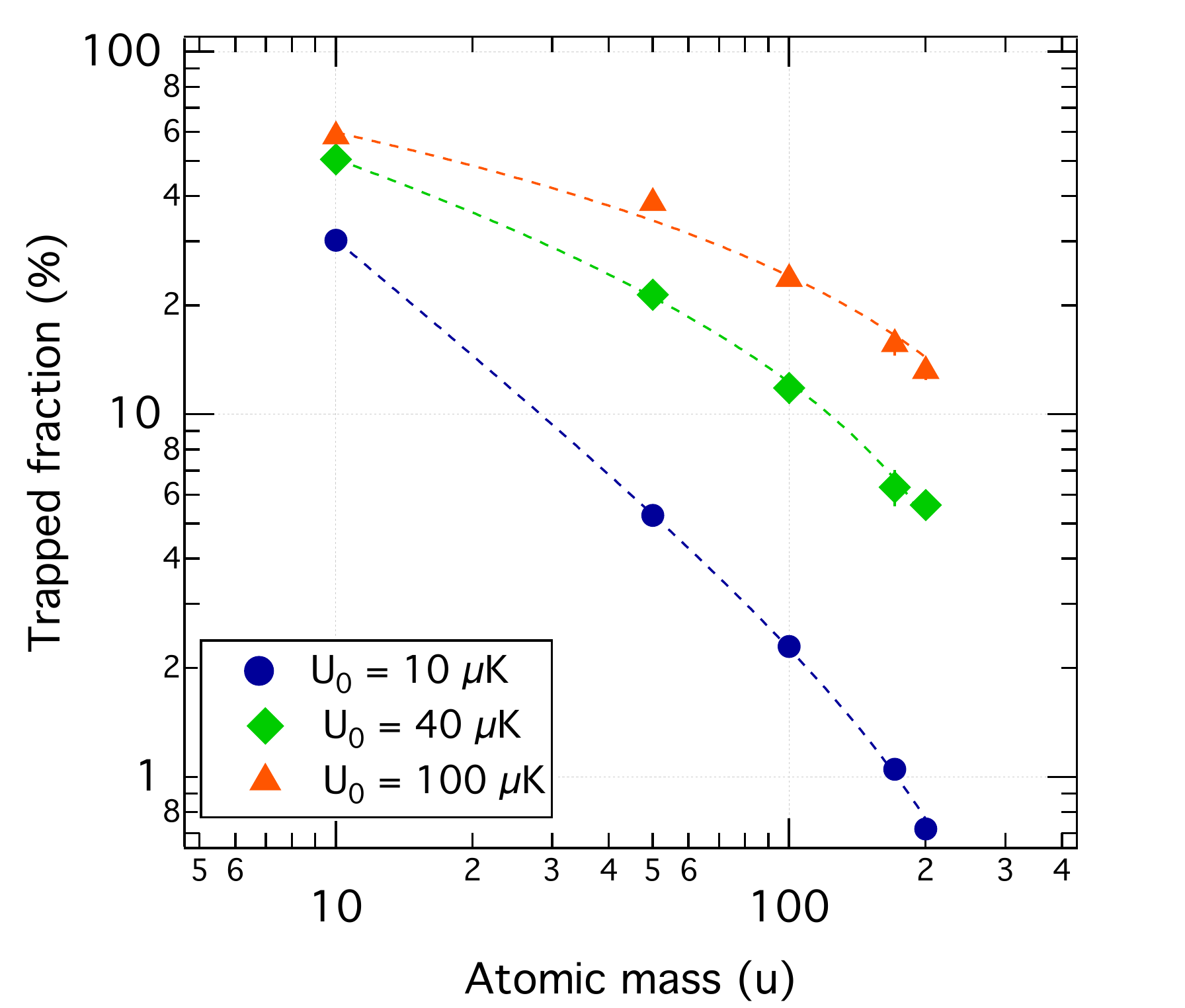}
        \caption{ Trapped fraction versus atomic mass at different lattice depths for $T_i=20$\,\uK.
         }
          \label{Fig_mass}
    \end{center}
\end{figure}

Fig.~\ref{Fig_mass} shows the trapped fraction versus atomic mass for  lattice depths of 10\,\uK\ (circles), 40\,\uK\  (diamonds) and 100\,\uK\ (triangles), and $\xi=0.5$. The initial cloud temperature was  20\,\uK. Lighter atoms are more easily trapped due to the fact that the action of the restoring force  $ \vec{\nabla} U(\vec{r}_{i}) /m$ is greater.  For the shallowest trap we have $f_t\propto m^{-1}$, as expected. At higher depths there is a roll-off, again due to $\xi<1$.   

In experiments where a large number of trapped atoms is important, such as in optical lattice clocks, it is advantageous to find the lattice beam waist size that produces the optimum transfer efficiency  for a given atomic cloud radius and optical power. 
As the waist size varies, both the transverse width of the trap and the trap depth change (seen in Eq.~\ref{EqDepth}), and both influence the trapped fraction.   For a given power, one may increase the waist size to better overlap the atom cloud, but this will incur a reduction in trap depth. In most cases it is not immediately apparent what the optimum waist size is.   

\begin{figure}[h!]
    \begin{center}
        \includegraphics[width=0.8\textwidth]{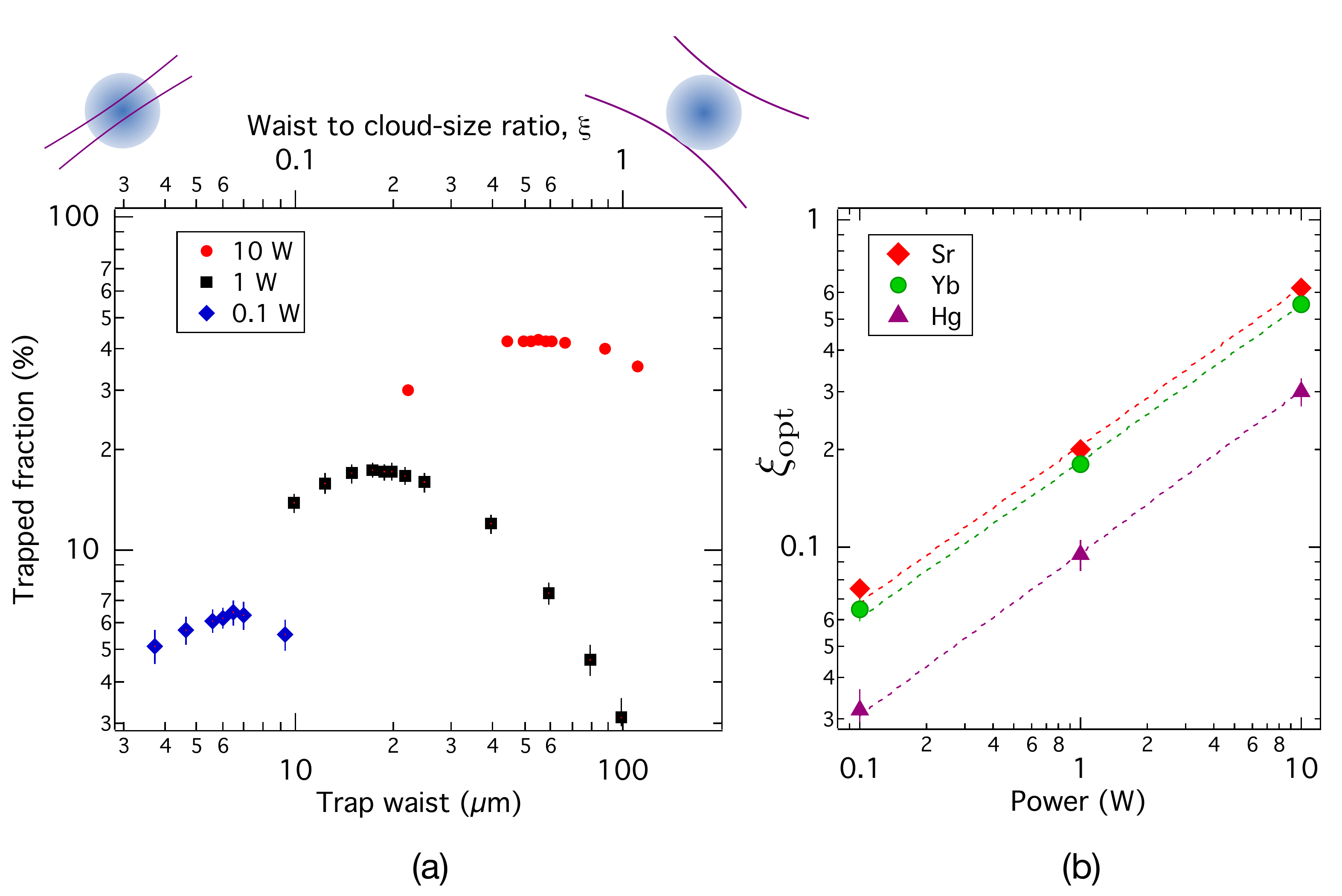}
        \caption{ (a) The fraction of ytterbium atoms  transferred to the lattice as a function of  laser waist size for three levels of laser power. The rms cloud radius was set to $100\,\mu$m.  The upper abscissa shows the trap waist normalised by the atomic cloud-size (i.e., $\xi$).  The sketches at the top are only symbolic  of the atom cloud and laser waist. 
        (b) Optimum $\xi$ versus optical power for Sr, Yb and Hg. 
        }
                \label{P_w0}  
    \end{center}
\end{figure}

We set  the rms cloud size  to be $100\,\mu$m with an initial temperature of $20\,\mu$K and calculate the lattice depth using the 
  dynamic polarizability of the Yb clock states at the magic wavelength, $\alpha_\mathrm{Yb}=3.07\times 10^{-39}$\,C\,m$^2\,$V$^{-1}$~\cite{Dzu2010}. 
   Fig.~\ref{P_w0}(a) shows the transfer fraction as a function of trap waist size for three different power levels, $P=0.1$\,W, 1\,W, and 10\,W.  We see there is an  optimum waist-to-cloud size ratio, $\xi_\mathrm{opt}$, for different levels of laser power.  At low power, a smaller waist is required to increase the trapping depth in order to retain some small fraction of atoms.  At high power, a greater fraction of atoms can be captured by better matching the lattice waist size to the atom cloud size. 
   Though not shown here, a simulation carried out with $\xi=4$ and $U_0/k_B T_i=200$ produces a 94\,\% trapping efficiency. 
   $\xi_\mathrm{opt}$ is plotted as a function of power in Fig.~\ref{P_w0}(b), where the data points for $^{171}$Yb are represented by circles. Simulations producing Fig.~\ref{P_w0}(a) have been repeated for Sr and Hg and  the  resultant $\xi_\mathrm{opt}$  values also plotted in  Fig.~\ref{P_w0}(b).
  The shift arises because of the different ac polarizabilities (at the respective magic wavelengths of 813.4\,nm and 362.6\,nm)~\cite{Por2008, Dzu2010}.  We assumed the values of $\alpha_\mathrm{Sr}=4.66\times 10^{-39}$\,C\,m$^2\,$V$^{-1}$ and $\alpha_\mathrm{Hg}=5.3\times 10^{-40}$\,C\,m$^2\,$V$^{-1}$.   The optimum $\xi$ ratio  scales almost with $\sqrt{P}$. The power law fits in Fig.~\ref{P_w0}(b) have  $\xi_\mathrm{opt}\propto P^{0.48}$.
  A change in initial cloud temperature causes a change in the trapped fraction, but does not affect  $\xi_\mathrm{opt}$.  This result can be applied generally with knowledge of the ac polarizability. 

\begin{figure}[h!]
    \begin{center}
        \includegraphics[width=0.5\textwidth]{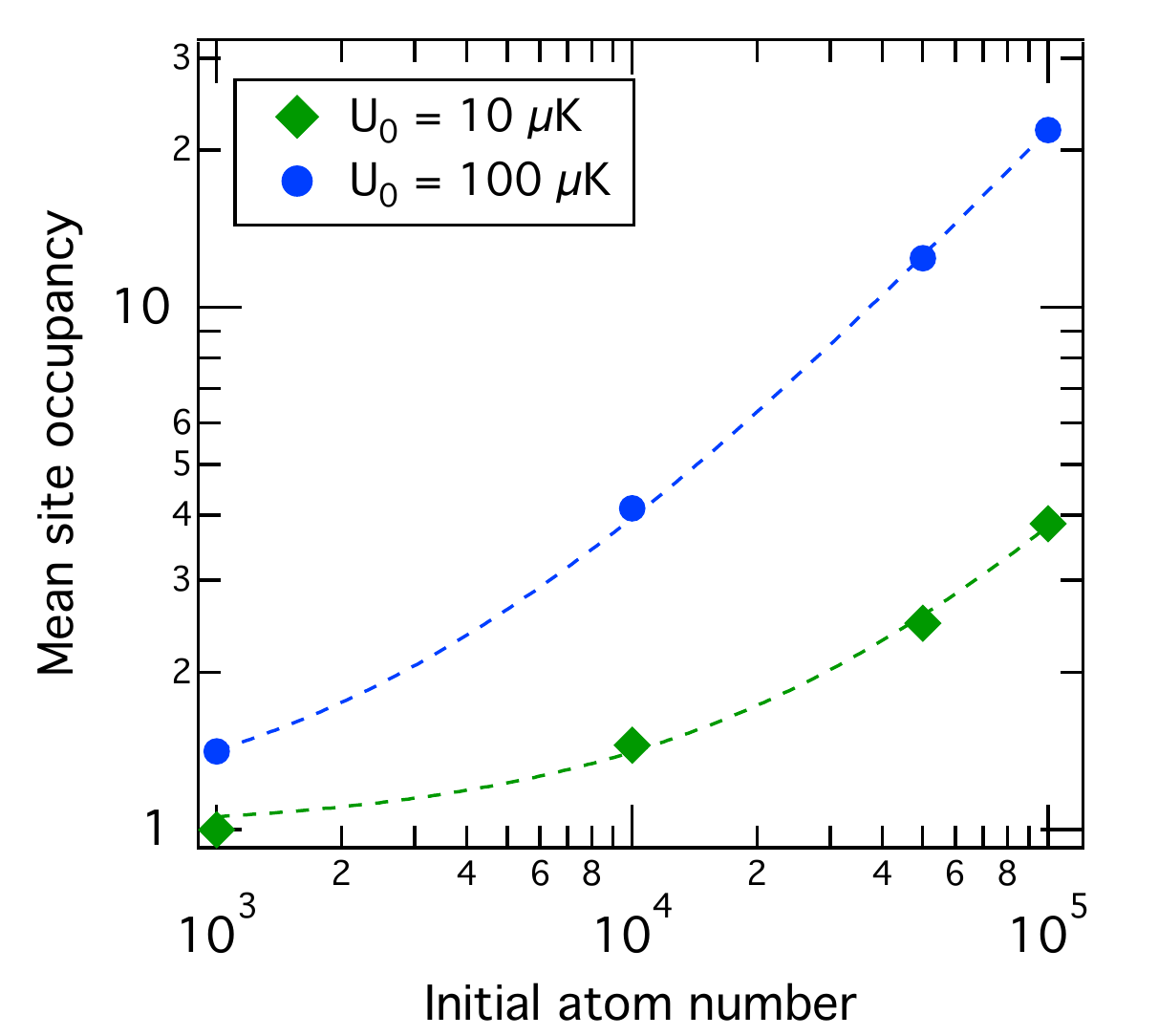}
        \caption{ 
         Mean site occupancy versus initial atom number for \Yb\ with trapping parameters: $w_0 = 80$\,\um, $r_a=120$\,\um, $\lambda=759.4$\,nm and $T_i=20$\,\uK.  
         }
          \label{FigOccup}
    \end{center}
\end{figure}

The number of atoms per site is also of interest, since collision or density shifts may impact experiments, such as in OLCs. 
In Fig.~\ref{FigOccup}, we plot the  mean site occupancy, $\bar{n}$, versus the initial  number or atoms, $N$, for two trap depths. The trapping parameters are: $w_0 = 80$\,\um, $r_a=120$\,\um, $\lambda=759.4$\,nm and $T_i=20$\,\uK. 
The dashed lines are  power law fits  of the form  $\bar{n}(N)=1+aN^x$, where $a$ is a constant, 
 In both cases the exponent is $x\approx0.84$.  We  consider  the mean over occupied sites only, hence $\bar{n}\geq 1$. 
In experiment, $10^5$ is a low number of atoms for a MOT cloud, but simulating a million atoms  takes a week of computational time on a standard pc.  Based on the progression here, for an initial $10^6$ atoms and $U_0=100$\,\uK, we may expect to have an average of $\sim150$ atoms per lattice site in  the same trapping conditions, with a corresponding number density of $\sim4\times10^{10}$\,cm$^{-3}$.

\section{Comparison with Experimental Data} 

\begin{figure}[h!]
    \begin{center}
        \includegraphics[width=0.6\textwidth]{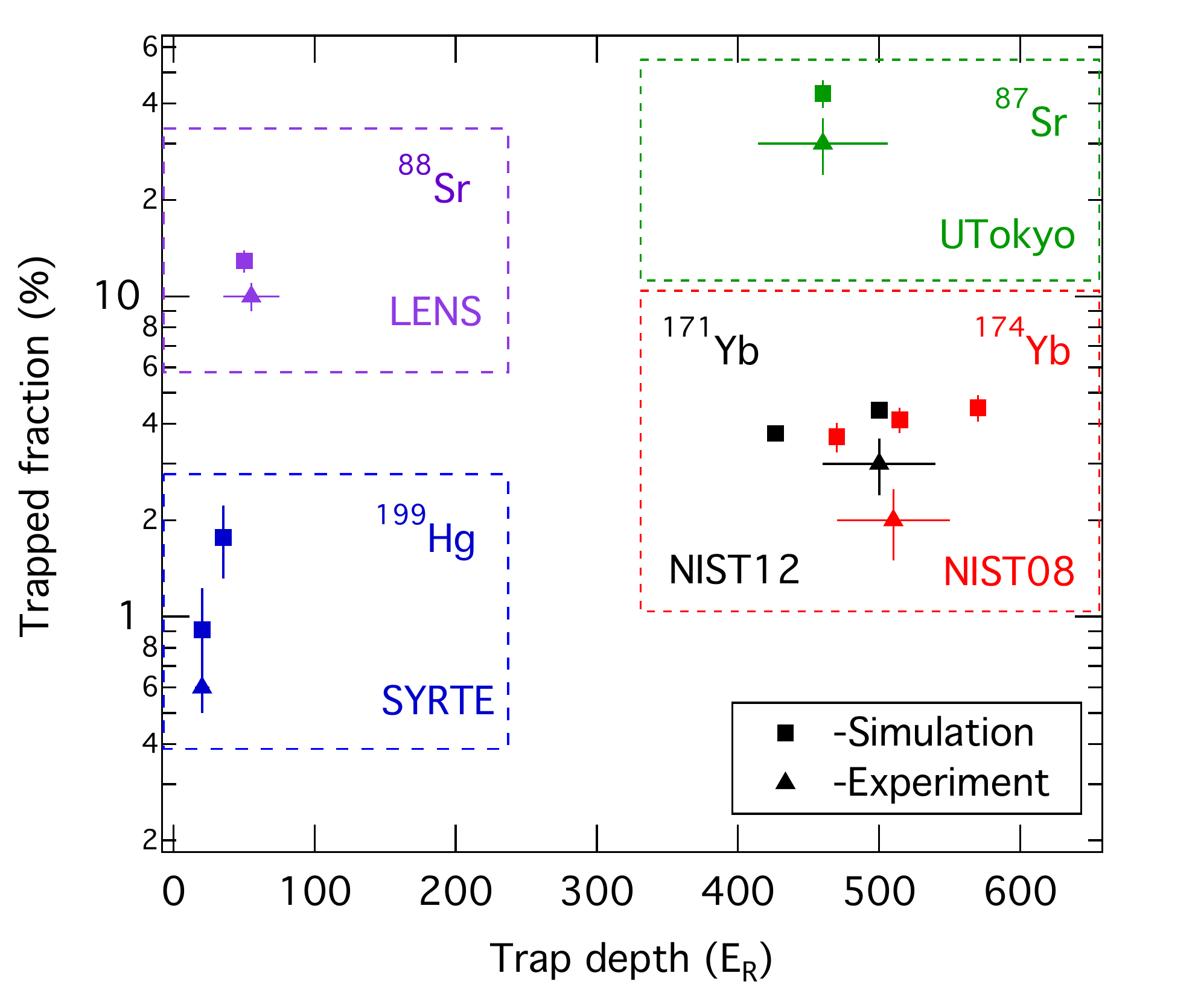}
        \caption{Comparison between simulation and experimental data for mercury (blue markers), strontium (purple and green markers), and ytterbium (black and red markers).  The square   (triangular) markers represent simulated   (experimental) values.  The labels represent the institutions where the experiments  were carried out. 
                }
                \label{FigExp}
    \end{center}
\end{figure}

Our simulations have been compared with a number of experimental results,  including those involving  atoms other than Yb.  Each experimental comparison is grouped by atomic species and considered in turn.   For comparisons with 
Yb we rely on data from NIST (USA), recorded in the works of Barber \textit{et al.}\cite{Bar2008a, Barber2003YtterbiumClock} (for  $^{174}$Yb) and Lemke \textit{et al.}\cite{Lem2009a,Lemke2012OpticalAtoms} (for  $^{171}$Yb). We  denote these results as  NIST08 and  NIST12, respectively. 
 With the parameters applicable to NIST08 (summarized in Table~\ref{SummaryTab}) 
  we compute a  transfer efficiency of 4.1\,(0.9)\%, which is higher than the  
 2\% transfer inferred  from \cite{Barber2003YtterbiumClock}.  
For NIST12  we evaluate a  trapped fraction of 4.5\,(0.8)\%, which again is slightly higher than  the 3\,\% transfer reported in \cite{Lemke2012OpticalAtoms}. The comparison is shown in Fig.~\ref{FigExp}, where we include computations for a number of different depths.   
Note, the 180\,\um\ cloud radius reported in \cite{Lemke2012OpticalAtoms} was used for both simulations.  One might expect that the cloud size was larger for $^{174}$Yb, thus reducing our estimated transfer fraction. 



   \begin{table} [] \caption{Summary of experimental parameters used for lattice loading simulations.} \label{SummaryTab}
   \vspace{-0.4cm} 
 \begin{tabular}{clclclclclclclclc}
   \vspace{-0.4cm}  \\
  \hline
 Label  & Atom  &  $U_0$ &  $U_0$  & $w_0\, (e^{-2})$  &  $r_a $ (rms) & $\xi$ &$T_\mathrm{MOT}$  \\ 
   (Refs.)      &         &  ($E_R$) &  (\uK) &  (\um) &  (\um) &   $w_0/r_a $& (\uK)           \\   
  \vspace{-0.5cm}  \\
 \hline
NIST08\cite{Bar2008a, Barber2003YtterbiumClock} & $^{174}$Yb   & 510 &  50 & 30 & 180 &  0.17&  40 \\
NIST12\cite{Lem2009a,Lemke2012OpticalAtoms} & $^{171}$Yb  & 500 & 50 & 45 & 180 & 0.25 &  25 \\
SYRTE\cite{McFerran2012Neutral-15} & $^{199}$Hg  & 20 &  7 & 120 & 110 & 1.1 &40 \\
LENS\cite{Pol2009,Tarallo2009DevelopmentClock} & $^{88}$Sr  &  55 &  9  & 30 &  100  &0.3 &1 \\
UTokyo\cite{Muk2003,Tak2003}  &  $^{87}$Sr &   460  &   76 &  34 & 60 &0.56 & 3  &\\
  \hline
 \end{tabular}
 \end{table}
 
A comparison is also made for the lattice trapping of $^{199}$Hg,  
 where atoms are trapped with laser radiation at $362.6$\,nm~\cite{McFerran2012Neutral-15}. The vertically oriented lattice laser (cavity enhanced) produced a potential depth of $U_0 \approx 20 \,E_R$.  The rms cloud radius was $\sim 110$\,\um\ and the lattice waist size 120\,\um.  The measured trapping efficiency was approximately $0.6\%$ for an atom cloud cooled to $\sim40\,\mu $K (blue triangle in Fig.~\ref{FigExp}).  Our simulation estimates  a trapped fraction of 0.9\,(0.3)\% for the same conditions, which is in fair agreement.  The large difference in trap depths between Yb and Hg arise primarily due to the different atomic polarizabilities ($\alpha_\mathrm{Yb}/\alpha_\mathrm{Hg}\sim5.8$).

 


 
 %

Two sets of experimental comparisons were made for strontium: one from LENS (Italy)  with $^{88}$Sr\cite{Pol2009,Tarallo2009DevelopmentClock} 
 and another  from  
 the University of Tokyo~\cite{Muk2003,Tak2003}.
  For $^{88}$Sr,  $\sim$10\% 
of the atoms were transferred to the optical lattice (purple triangle in Fig.~\ref{FigExp}).
The higher transfer efficiency compared with the trapping of Yb or Hg is due principally to the  lower cloud temperatures achieved by the use of narrow-line laser cooling  (where there is a Doppler cooling limit of \si0.2\uK). 
  Strontium is also aided by its smaller mass.
   The rms cloud radius, $r_a$, has been deduced from the quoted densities and atomic populations~\cite{Pol2009,Tarallo2009DevelopmentClock}. 
 For  the LENS comparison we estimate a trapped percentage of 13\,(1)\,\% at a depth of $55\,E_R$, which slightly above the experimental value.   The Tokyo experiment had a  deeper trap producing a transfer fraction of $\sim30$\,\%. Based on the reported experimental parameters, our simulation predicts a 43(3)\,\% transfer.   A comparison may also be made against experimental data from JILA~\cite{Ludlow2008thesis,Cam2008,Bla2009}, however, there a lower  transfer efficiency may have been deliberately employed to minimise density dependent shifts in the clock transition~\cite{Cam2009}. 

\section{Conclusions}

Through simulation we have investigated the loading efficiency of atoms into an optical lattice trap with light that is far detuned from any atomic resonance. The atoms are loaded from a spherically symmetric normal distribution  of cold atoms of specified size and temperature.   Different atomic species are considered by inputting the appropriate atomic polarizability and trapping wavelength.
The trap waist and atomic cloud size geometries have been set to reflect  experimental conditions, and  
the  lattice trap depth is varied by selection of laser beam power or  directly in terms of  energy. 
The primary outputs of the simulation are:  the fraction of atoms transferred to the lattice, the resultant velocity distribution, the temperature of the trapped atoms, and the site occupancy.
These outputs are a useful guide when designing ultracold atom experiments requiring  lattice traps.  Although our focus has been on Yb, Hg and Sr at magic wavelengths, the simulated outputs can be applied to other atomic species by use of appropriate scalings. 

   The simulations have been used to evaluate the optimum ratio of  lattice waist size to rms atom cloud radius for a fixed optical power, granting maxim transfer efficiency. 
This is applicable in situations where a set amount of power is available and   one has freedom to adjust the laser waist size, or cloud size, for optimum transfer. 
 The optimum trap waist  to cloud size ratio scales approximately with square root of the lattice power.  Our result can be applied to an arbitrary atomic species after 
  accounting for the change in ac polarizabilty.

 The simulation of trapping efficiency has been compared against experimental results in Yb, Hg and Sr,
  showing relatively good agreement, but with some over-estimation of trapped fractions, likely due
 to the idealisation of the trap structure. 
 One could treat our estimates as an upper limit for the expected transfer efficiency, as we assume optimum alignment between the lattice beam and atom cloud, and no losses due to collisional processes. 
    An extension of the simulations would be to include simultaneous operation of the lattice trap and \MOT, as this will affect the distribution of atoms prior to the MOT fields being removed. A further extension may be to consider 
     non-normal distributions at high atomic density~\cite{Wal1990}, or elongated distributions.  
    The Matlab code with graphical user interface and User Documentation  are available in   the Supplementary Materials.  The software package can also simulate a single focused beam trap or crossed beam trap. 
     
\section*{Acknowledgements}
We thank P. Atkinson for assisting with the preparation of the manuscript and J. Schelfhout for insightful comments and careful scrutiny of the paper.  R.W. is grateful to the Australian Research Council's Centre of Excellence for Engineered Quantum Systems for travel support. 

\section*{Disclosures}  The authors declare no conflicts of interest.



\end{document}